\RequirePackage{snapshot} 
\documentclass[preprint2]{aastex}
\usepackage{natbib}
\usepackage{graphicx}
\usepackage{natbib}
\usepackage{ctable}
\usepackage{placeins}
\usepackage[nolist, nostyles]{glossaries}
\glsdisablehyper

\graphicspath{{./plots/}, {}}

\usepackage{xspace}

\newcommand{\msun}{\hbox{$M_{\odot}$}}

\newglossaryentry{vrad}{name={radial velocity~}, text={radial velocity}, symbol={\ensuremath{v_\textrm{rad}}}, description={radial velocity}, sort=vrad}
\newglossaryentry{vrot}{name={stellar rotation~}, name={stellar rotation}, symbol={\ensuremath{v_\textrm{rot}}}, description={radial velocity}, sort=vrot}

\newcommand{\xray}{X-ray}

\newcommand{\nucl}[2]{\ensuremath{^{#2}\textrm{#1}}}
\newglossaryentry{angstrom}{name=\AA, description={unit of length $10^{-10}$\,m}, sort=angstrom}
\newglossaryentry{nir}{name=NIR,description={near infrared},first = {near infrared (NIR)}}
\newglossaryentry{psf}{name=PSF,description={point-spread function},first = {point-spread function (PSF)}}
\newglossaryentry{fwhm}{name=FWHM,description={Full Width Half Maximum},first = {FWHM}}
\newglossaryentry{rms}{name=RMS,description={Root Mean Square},first = {RMS}}
\newglossaryentry{signalnoise}{name=S/N,description={signal to noise}}
\newglossaryentry{uv}{name=UV,description={ultra violet},first = {ultra violet (UV)}}
\newglossaryentry{halpha}{name=\ensuremath{\textrm{H}\alpha}, description={First line of the Balmer series at 6563\,\AA}, sort=halpha}
\newglossaryentry{mgb}{name={Mg \textsc{i} b}, description={Triplet at 5167\,\AA, 5173\,\AA and 5184\,\AA}}
\newglossaryentry{sobolevapprox}{name={Sobolev approximation}, description={Lines are approximation with an infinitley thin interaction region \citep[e.g. no broadening][]{1960mes..book.....S}}, first={Sobolev approximation }}
\newglossaryentry{radeq}{name={radiative equilibrium}, description={The net flux of energy between matter and radiation field is zero}}
\newglossaryentry{nebularapprox}{name={nebular approximation}, description={Assumes that the plasma condition are controlled by a central radiation source. The radiation field decreases with the distance to the source by geometrical dilution. See \citet{1978stat.book.....M} for details}}
\newglossaryentry{modnebularapprox}{name={modified nebular approximation}, description={In contrast to \gls{nebularapprox} where only geometrical dilution is taken into account, the modified nebular approximation also takes dilution by other radiative processes into account }, first={modified nebular approximation}, parent=nebularapprox}
\newglossaryentry{thompsonscat}{name={Thomson scattering}, description={Scattering of photons on low energy electrons}}
\newglossaryentry{lte}{name={LTE}, description={Local Thermodynamic Equilibrium}, first={local thermodynamic equilibrium (LTE)}}
\newglossaryentry{lsr}{name={LSR}, description={Local Standard of Rest}, first={\textit{local standard of rest} (LSR)}}
\newglossaryentry{mc}{name={MC}, description={Monte Carlo}, first={\textit{Monte Carlo} (MC)}}
\newglossaryentry{wcs}{name={WCS}, description={world coordinate system}, first={world coordinate system (WCS)}}
\newglossaryentry{cmf}{name=CMF, text=CMF, first=Comoving Frame (CMF henceforth), description={Comoving Frame}}

\newglossaryentry{sfit}{name=SFIT, text=\textsc{sfit}, description={spectral fitting program for hot stars \citep{2001A&A...376..497J}}, first={\textsc{sfit} \citep{2001A&A...376..497J}}}
\newglossaryentry{iraf}{name=IRAF, text=\textsc{iraf}, description={Image Reduction and Analysis Facility maintained by NOAO}, first={\textsc{iraf}\protect\footnote{IRAF: the Image Reduction and Analysis Facility is distributed by the National Optical Astronomy Observatory, which is operated by the Association of Universities for Research in Astronomy (AURA) under cooperative agreement with the National Science Foundation (NSF).}}}
\newglossaryentry{pyraf}{name=PyRAF, text=\textsc{PyRAF}, description={Python wrap of \gls{iraf} maintained by STSCI}, first=\textsc{PyRAF} \protect\footnote{PyRAF is a product of the Space Telescope Science Institute, which is operated by AURA for NASA.}}
\newglossaryentry{scipy}{name=SciPy, text=\textsc{Scipy}, description={Scientific Python \cite{Jones:2001fk}}}
\newglossaryentry{moog}{name=MOOG,text={\textsc{moog}}, description={spectral synthesis software \citep{1973ApJ...184..839S}}, first={\textsc{Moog} \citep{1973ApJ...184..839S}}}
\newglossaryentry{atlas9}{name=ATLAS9,description={grid of stellar atmospheres \citep{2004astro.ph..5087C}}, first={ATLAS9 \citep{2004astro.ph..5087C}}}
\newglossaryentry{vald}{name=VALD,description={Vienna Atomic Line Database \citep{2000BaltA...9..590K}}, first={Vienna Atomic Line Database \citep[VALD;][]{2000BaltA...9..590K}}}
\newglossaryentry{sextractor}{name=SExtractor, text=\textsc{SExtractor}, description={Source Extractor photometry program \citep{1996A&AS..117..393B}}, first={\textsc{SExtractor} \citep{1996A&AS..117..393B}}}

\newglossaryentry{swarp}{name=SWarp, text=\textsc{SWarp}, description={SWarp \citep{2002ASPC..281..228B}}, first={\textsc{SWarp} \citep{2002ASPC..281..228B}}}

\newglossaryentry{idl}{name=IDL,text={\textsc{idl}}, description={Interactive Data Language}}
\newglossaryentry{makee}{name=MAKEE,text=\textsc{makee}, description={MAuna Kea Echelle Extraction by Tom Barlow available}}% at \verb+http://spider.ipac.caltech.edu/staff/tab/makee/index.html+}}
\newglossaryentry{minuit}{name=MINUIT,text={\textsc{minuit}}, description={collection of numerical optimization tools \citep{James:1975dr}}}
\newglossaryentry{migrad}{name=MIGRAD,text={\textsc{migrad}}, description={numerical gradient optimization tools - part of \gls{minuit}}}
\newglossaryentry{dolphot}{name=DOLPHOT, text=DOLPHOT, description=photometry package for HST, first=DOLPHOT \citep{2000PASP..112.1383D}}
\newglossaryentry{synphot}{name=synphot, text={\textsc{synphot}}, description={synthetic photometry package from STSCI}, first={\textsc{synphot}\protect\footnote{\textsc{synphot} is a product of the Space Telescope Science Institute, which is operated by AURA for NASA.}}}
\newglossaryentry{chianti}{name=CHIANTI, text=CHIANTI, description= CHIANTI Database 7.1, first =CHIANTI 7.1 \citep{1997A&AS..125..149D,2012ApJ...744...99L}}
\newglossaryentry{synpp}{name=SYNPP, text=SYN++, description= SYN++ software, first =SYN++ \citep{2011PASP..123..237T}}
\newglossaryentry{tardis}{name=TARDIS, text=TARDIS, description= TARDIS MC code, first = Temporal And Radiative Diffusion In Supernovae (TARDIS)}
\newglossaryentry{astropy}{name=ASTROPY, text=\textsc{astropy}, description=\textsc{astropy} framework, first = \textsc{astropy} \citep{2013A&A...558A..33A}}

\newglossaryentry{2mass}{name=2MASS,description={Two Micron All Sky Survey \citep{2006AJ....131.1163S}}, first={Two Micron All Sky Survey \citep{2006AJ....131.1163S}}}
\newglossaryentry{nomad}{name=NOMAD,first={Naval Observatory Merged Astrometric Dataset \citep[NOMAD; ][]{2005yCat.1297....0Z}}, description={Naval Observatory Merged Astrometric Dataset}}
\newglossaryentry{wifes}{name=WIFES, text=\textsc{WiFeS}, first={\textsc{WiFeS} \citep{2007Ap&SS.310..255D}},  description={Wide Field Spectrograph - \gls{ifu} mounted on the 2.3\,m telescope at Siding Spring Observatory}}
\newglossaryentry{scp}{name=SCP,description={Supernova Cosmology Project, led by Saul Perlmutter}, first={Supernova Cosmology Project (SCP)}}
\newglossaryentry{hzsns}{name=HZSNS,description={High Z Supernova Search, led by Brian Schmidt}, first={High Z Supernova Search (HZSNS)}}
\newglossaryentry{vlt}{name=VLT,description={Very Large Telescope located on Cerro Paranal (Chile)}, first={Very Large Telescope (VLT)}}
\newglossaryentry{flames}{name=FLAMES,description={Multi-object, intermediate and high resolution spectrograph mounted on the  \gls{vlt}}}
\newglossaryentry{hires}{name=HIRES, description={High Resolution Echelle Spectrometer mounted on the Keck Telescope}, first={High Resolution Echelle Spectrometer \citep[HIRES;][]{1994SPIE.2198..362V}}}
\newglossaryentry{lris}{name=LRIS,description={Low Resolution Imaging Spectrometer mounted on the Keck Telescope}, first={Low-Resolution Imaging Spectrometer \citep[LRIS;][]{Oke95}}}

\newglossaryentry{essence}{name=ESSENCE,description={The `Equation of State: SupErNovae trace Cosmic Expansion' project \citep[ESSENCE;][]{2002AAS...201.7809G}}, first={`The Equation of State: SupErNovae trace Cosmic Expansion' \citep[ESSENCE;][]{2002AAS...201.7809G}}}
\newglossaryentry{ifu}{name=IFU,description={Optical instrument combining spectrographic and imaging capabilities, used to obtain spatially resolved spectra}, first={Integral Field Unit (IFU)}, firstplural={Integral Field Units (IFUs)}} 

\newglossaryentry{besancon}{name=Besan\c{c}on Model, description={Model of stellar population synthesis of the Galaxy, including kinematics.}}%  \verb+http://model.obs-besancon.fr+} }, nonumberlist=true}

\newglossaryentry{int}{name=INT,description={Isaac Newton 2.5\,m Telescope}, first={Isaac Newton 2.5\,m Telescope (INT)}}
\newglossaryentry{iau}{name=IAU,description={International Astronomical Union}, first={IAU}}
\newglossaryentry{chandra}{name=Chandra,description={Chandra \xray\ Observatory (space-based)}}
\newglossaryentry{hst}{name=HST,description={Hubble Space Telescope}}
\newglossaryentry{wfpc2}{name=WFPC2,description={Wide-Field Planetary Camera 2 mounted on the \gls{hst}}, first={Wide-Field Planetary Camera 2 (WFPC2)}}
\newglossaryentry{acs}{name=ACS,description={Advanced Camera for Surveys mounted on the \gls{hst}}, first={Advanced Camera for Surveys (ACS)}}
\newglossaryentry{snls}{name=SNLS,description={Supernova Legacy Survey \citep{2003AAS...203.8209P}}, first={Supernova Legacy Survey \citep[SNLS;][]{2003AAS...203.8209P}}}
\newglossaryentry{dass}{name=DASS, description={Digitized Astronomy Supernova Survey \citep{1975PASP...87..565C}}, first={Digitized Astronomy Supernova Survey \citep[DASS;][]{1975PASP...87..565C}}}
\newglossaryentry{bait}{name=BAIT, description={Berkley Automatic Imaging Telescope \citep{1993PASP..105.1164R}}, first={Berkley Automatic Imaging Telescope \citep[BAIT;][]{1993PASP..105.1164R}}}
\newglossaryentry{kait}{name=KAIT, description={Katzman Automatic Imaging Telescope \citep{2001ASPC..246..121F}}, first={Katzman Automatic Imaging Telescope \citep[KAIT;][]{2001ASPC..246..121F}}}
\newglossaryentry{loss}{name=LOSS, description={Lick Observatory Supernova Search  \citep{2000AIPC..522..103L}}, first={Lick Observatory Supernova Search \citep[LOSS;][]{2000AIPC..522..103L}}}
\newglossaryentry{ctss}{name=CTSS,description={Cal\'{a}n/Tololo Supernova Survey \citep{1993AJ....106.2392H}}, first={Cal\'{a}n/Tololo supernova survey \citep[CTSS;][]{1993AJ....106.2392H}}}
\newglossaryentry{ctio}{name= CTIO, description={Cerro Tololo Inter-American Observatory}, first={Cerro Tololo Inter-American Observatory (CTIO)}}
\newglossaryentry{ptf}{name=PTF, description={Palomar Transient Factory \citep{2009PASP..121.1334R}}, first={Palomar Transient Factory \cite[PTF;][]{2009PASP..121.1334R}}}
\newglossaryentry{batse}{name=BATSE, description={Burst and Transient Source Experiment mounted on the Compton Gamma Ray Observatory}, first={Burst and Transient Source Experiment (BATSE)}}
\newglossaryentry{bepposax}{name=BeppoSAX, description={\xray\ satellite named in honor of Giuseppe "Beppo" Occhialini}}
\newglossaryentry{rosat}{name=ROSAT, description={short for R\"{o}ntgensatellit}, first={ROSAT}}
\newglossaryentry{hete2}{name=HETE2, description={High Energy Transient Explorer}, first={High Energy Transient Explorer (HETE)}}

\newglossaryentry{gnirs}{name=GNIRS, description={Gemini Near InfraRed Spectrograph mounted on the Gemini North Telescope}}
\newglossaryentry{gmosn}{name=GMOS, description={Gemini Multi Object Spectrograph mounted on the
 Gemini North Telescope}, first={GMOS \citep[Gemini Multi Object Spectrograph;][]{2004PASP..116..425H}}}
\newglossaryentry{swift}{name=Swift, description={Swift Gamma-Ray Burst Mission}}
\newglossaryentry{vla}{name=VLA, description={Very Large Array radio telescope located in North America}, first={Very Large Array (VLA)}}
\newglossaryentry{evla}{name=EVLA, description={Extended Very Large Array radio telescope located in North America}, first={Extended Very Large Array (EVLA)}}
\newglossaryentry{sdss}{name=SDSS, description={Sloan Digital Sky Survey}}
\newglossaryentry{dss}{name=DSS, description={Digitized Sky Survey}}
\newglossaryentry{skymapper}{name=SkyMapper, description={SkyMapper telescope \citep{2007PASA...24....1K}}, first={SkyMapper \citep{2007PASA...24....1K}}}
\newglossaryentry{panstarrs}{name=PanSTARRS, description={Panoramic Survey Telescope \& Rapid Response System \citep{2004SPIE.5489...11K}}, first={Panoramic Survey Telescope \& Rapid Response System \citep[PanSTARRS;][]{2004SPIE.5489...11K}}}
\newglossaryentry{lsst}{name=LSST, description={Large Synoptic Survey Telescope}, first={Large Synoptic Survey Telescope \citep[LSST;][]{2006AAS...209.8604P}}}
\newglossaryentry{ppmxl}{name=PPMXL, description={PPMXL Catalog of Positions and Proper Motions on the ICRS \citep{2010AJ....139.2440R}}}
\newglossaryentry{gaia}{name=GAIA, description={Global Astrometric Interferometer for Astrophysics \citep{2001A&A...369..339P}}, first={Global Astrometric Interferometer for Astrophysics \citep[GAIA;][]{2001A&A...369..339P}}}
\newglossaryentry{ligo}{name=LIGO, description={Laser Interferometer Gravitational Wave Observatory}, first={Laser Interferometer Gravitational Wave Observatory \citep[LIGO;][]{1992Sci...256..325A}}}
\newglossaryentry{aligo}{name=Advanced LIGO, description={Advanced LIGO}, sort=ligo2}
\newglossaryentry{lisa}{name=LISA, description={Laser Interferometer Space Antenna \citep{1994ESAJ...18..219J}}, first={Laser Interferometer Space Antenna \citep[LISA;][]{1994ESAJ...18..219J}}}
\newglossaryentry{ccd}{name=CCD,description={Charged Coupled Device}, first={charged coupled device (CCD)}, firstplural={charged coupled devices (CCDs)}}

\newcommand{\sn}[2]{SN~#1#2\xspace}

\newglossaryentry{irc}{name=IRC, text={IRC}, description={infrared catastrophe}, first={infrared catastrophe \citep[IRC;][]{1980PhDT.........1A}}}

\newglossaryentry{sn}{name=Supernova, text={SN}, plural={SNe}, description={exploding star}, nonumberlist=true, first={supernova (SN)}, firstplural={supernovae (SNe)}}
\newglossaryentry{snia}{name=Type~Ia (SN~Ia), text={SN~Ia}, description={Thermonuclear explosion of a white dwarf - spectra show no hydrogen but a strong silicon line},first={Type~Ia supernova (SN~Ia)}, firstplural={Type Ia supernovae (SNe~Ia)}, plural={SNe~Ia}, parent=sn, nonumberlist=true}
\newcommand{\sneia}{\glspl*{snia}\xspace}
\newcommand{\snia}{\gls*{snia}\xspace}

\newglossaryentry{branchnormal}{name={branch-normal}, text=\textit{Branch-normal}, description={Large homogeneous class of Type Ia Supernovae, defined in \citet{1993AJ....106.2383B}}, first={\textit{Branch-normal} SNe Ia \citep{1993AJ....106.2383B}}, parent=snia} 
\newglossaryentry{91t}{name={91T-like}, description={Luminous class of Type Ia supernovae similar to \sn{1991}{T} \citep{1992AJ....103.1632P}} , first={91T-like}, parent=snia} 
\newglossaryentry{91bg}{name={91bg-like}, description={Faint class of Type Ia supernovae similar to \sn{1991}{bg} \citep{1992AJ....104.1543F}}, first={91bg-like}, parent=snia} 
\newglossaryentry{02cx}{name={02cx-like}, description={Peculiar class of Type Ia supernovae similar to \sn{2002}{cx} \citep{2003PASP..115..453L}}, first={02cx-like \sneia\ \citep{2003PASP..115..453L}}, parent=snia} 

\newglossaryentry{snibc}{name=Type~Ib/c, text={SN~Ib/c}, description={Collapse of the core of a massive star -  spectrum shows no hydrogen and no silicon line},first={Type~Ib/c supernova (SN~Ib/c)}, firstplural={Type~Ib/c supernovae (SNe~Ib/c)}, plural={SNe~Ib/c}, parent=sn}

\newglossaryentry{snib}{name=Type~Ib, text={SN~Ib}, description={Spectrum shows no hydrogen and no silicon, but helium line},first={Type Ib supernova (SN~Ib)}, firstplural={Type~Ib supernovae (SNe~Ib)}, plural={SNe~Ib}, parent=snibc}

\newglossaryentry{snic}{name=Type~Ic, text={SN~Ic}, description={Spectrum shows no hydrogen, no silicon and no helium line},first={Type~Ic supernova (SN~Ic)}, firstplural={Type~Ic supernovae (SNe~Ic)}, plural={SNe~Ic}, parent=snibc}

\newglossaryentry{snii}{name=Type~II, text={SN~II}, description={Collapse of the core of a massive star - spectrum shows strong hydrogen line},first={Type~II supernova (SN~II)}, firstplural={Type~II supernovae (SNe~II)}, plural={SNe~II}, parent=sn}

\newglossaryentry{sniib}{name=Type~IIb, text={SN~IIb}, description={Spectrum shows hydrogen and helium lines},first={Type~IIb supernova (SN~IIb)}, firstplural={Type~IIb supernovae (SNe~IIb)}, plural={SNe~IIb}, parent=snii}

\newglossaryentry{sniip}{name=Type~II~Plateau (Type IIP), text={SN~IIP}, description={Lightcurve shows plateau},first={Type~IIP supernova (SN~IIP)}, firstplural={Type~II Plateau supernovae \citep[SNe~IIP;][]{1979A&A....72..287B}}, plural={SNe~IIP}, parent=snii}

\newglossaryentry{sniil}{name=SN~II~Linear, text={SN~IIL}, description={Lightcurve shows no plateau, but linear decline},first={Type~IIL supernova (SN~IIL)}, firstplural={Type~II~Linear supernovae \citep[SNe~IIL;][]{1990MNRAS.244..269S}}, plural={SNe~IIL}, parent=snii}

\newglossaryentry{sniin}{name=Type II narrow-lined (Type IIn), description={Spectrum shows narrow lines},first={Type~II~narrow-lined supernova (SN IIn)}, firstplural={Type~IIn supernovae (SNe~IIn)}, plural={SNe~IIn}, parent=snii}

\newglossaryentry{snr}{name=Remnant (SNR), text=SNR, description={Remnant left visible post-explosion}, first={supernova remnant (SNR)}, firstplural={supernova remnants (SNRs)}, parent=sn}

\newglossaryentry{dtd}{name=DTD,description={delay time distribution - expected supernova rate over time after a brief outburst of starformation},first={delay time distribution (DTD)}, firstplural={delay time distributions (DTDs)}, plural=DTDs}

\newglossaryentry{hvg}{name=HVG,description={high velocity gradient - Type Ia supernovae with a fast evolution of photospheric velocity},first={high velocity group (HVG)}, firstplural={high velocity groups (HVGs)}, plural=HVGs, parent=snia}

\newglossaryentry{lvg}{name=LVG,description={low velocity gradient - Type Ia supernovae with a slow evolution of photospheric velocity},first={low velocity group (LVG)}, firstplural={low velocity groups (LVGs)}, plural=LVGs, parent=snia}

\newglossaryentry{wd}{name=white dwarf (WD), text=WD, description={White Dwarf - extremely dense stellar remnant}, first={white dwarf (WD)}}
\newglossaryentry{onemgwd}{name= Oxygen/Neon (ONe), text={ONe-WD},description={Oxygen/Neon White Dwarf}, first={oxygen/neon White Dwarf (ONe-WD)}, parent=wd}
\newglossaryentry{cowd}{name=carbon/oxygen (CO), text={CO-WD}, description={carbon/oxygen white dwarf}, first={carbon/oxygen white dwarf (CO-WD)}, firstplural = {carbon/oxygen white dwarfs (CO-WDs)}, parent=wd}

\newglossaryentry{sds}{name=SD-Scenario,description={single-degenerate scenario (single white dwarf accreting from non-degenerate companion)}, first={single-degenerate scenario (SD-scenario)}}

\newglossaryentry{dds}{name=DD-Scenario, description={double degenerate scenario (merging of two white dwarfs)}, first={double-degenerate scenario (DD-scenario)}}

\newglossaryentry{sss}{name=SSS, text={supersoft \xray\ source}, description={supersoft \xray\ source - believed to be emitted by nuclear fusion on a white dwarf's surface}}%, first={supersoft \xray\ source (SSS)}, firstplural={supersoft \xray\ sources (SSS)}}

\newglossaryentry{amcvn}{name=AM CVn, description={AM Canum Venaticorum star (white dwarf accreting hydrogen poor matter from a companion star; see \cite{2005ASPC..330...27N})}}

\newglossaryentry{rlof}{name=RLOF, description={Roche Lobe Overflow (see \citet{1971ARA&A...9..183P} for a more detailed description)}, first={Roche-lobe overflow (RLOF)}}

\newglossaryentry{mchan}{name={Chandrasekhar mass~}, text={Chandrasekhar~mass}, symbol={\ensuremath{M_\textrm{Chan}}}, plural={Chandrasekhar~masses}, description={Mass when the core of a star collapses due to insufficient degeneracy pressure - for a white dwarf $\approx1.38\,M_\odot$ see \citet{1931ApJ....74...81C}}, first={Chandrasekhar~mass \citep[$M_\textrm{Chan}=1.38\,M_\odot$;][]{1931ApJ....74...81C}}, sort=mchan}

\newglossaryentry{w7}{name={W7 model},description={W7 model \citep{1984ApJ...286..644N}},first = {W7 model \citep{1984ApJ...286..644N}}}

\newglossaryentry{ew}{name=Equivalent Width, text={EW}, description={width of a rectangle that has the same area as a spectral line when taken to zero flux}, first={equivalent width (EW)}, firstplural={equivalent widths (EWs)}}
\newglossaryentry{agb}{name=AGB,description={Asymptotic Giant Branch}}
\newglossaryentry{cmb}{name=CMB,description={Cosmic Microwave Background}}
\newglossaryentry{csm}{name=CSM,description={Circumstellar Medium}, first={circumstellar medium (CSM)}}
\newglossaryentry{csi}{name=CSI,description={Circumstellar Interaction}, first={circumstellar interaction (CSI)}}
\newglossaryentry{ism}{name=ISM,description={Interstellar Medium}, first={interstellar medium (ISM)}}
\newglossaryentry{ige}{name=IGE,description={Iron Group Element}, first={iron group element (IGE)}, firstplural={iron group elements (IGEs)}}
\newglossaryentry{epm}{name=EPM,description={Expanding Photosphere Method \citep{1974ApJ...193...27K}}, first={Expanding Photosphere Method (EPM)}}
\newglossaryentry{aic}{name=AIC,description={Accretion Induced Collapse}, first={accretion induced collapse (AIC)}}
\newglossaryentry{ime}{name=IME,description={Intermediate Mass Element}, first={intermediate mass element (IME)}, firstplural={intermediate mass elements (IMEs)}}
\newglossaryentry{h0}{name=\ensuremath{H_0},description={Hubbles constant}}
\newglossaryentry{nse}{name=NSE,description={Nuclear Statistical Equilibrium}, first={nuclear statistical equilibrium (NSE)}}
\newglossaryentry{cdm}{name=CDM,description={Cold Dark Matter}}
\newglossaryentry{grb}{name=GRB,description={Gamma Ray Burst}, first={Gamma Ray Burst (GRB)}, firstplural={Gamma Ray Bursts (GRBs)}}
\newglossaryentry{donor}{name=donor,description={non-degenerate companion in the \gls{sds}}}
\newglossaryentry{mainsequence}{name=main sequence,description={main sequence star}}
\newglossaryentry{redgiant}{name=red giant,description={red giant star}}
\newglossaryentry{mlcs}{name=MLCS,description={Multicolor Light Curve Shape method \citep[MLCS;][]{1996ApJ...473...88R}}, first={Multicolor Light-Curve Shape method \citep[MLCS;][]{1996ApJ...473...88R}}}
\newglossaryentry{rsoph}{name=RS~Ophiuci ,description={white dwarf accreting from a red giant - assumed progenitor of the \gls{sds}}, sort=rsoph}
\newglossaryentry{usco}{name=U~Scorpii,description={white dwarf accreting from a main sequence star - assumed progenitor of the \gls{sds}}, sort=usco}
\newglossaryentry{rcw86}{name=RCW~86,description={supernova remnant sometimes associated with \sn{185}{}}, sort=rcw86}
\newglossaryentry{casa}{name=Cas~A,description={Cassiopeia A supernova remnant - probably a \gls{snib} event}}
\newglossaryentry{cepheid}{name=Cepheid,description={very luminous variable star with a strong luminosity period relationship}}
\newglossaryentry{urca}{name=Urca, text=\textit{Urca}, description={process predominatly contributing to cooling in stars. The \textit{Urca} process consists of alternating electron-capture and $\beta^{-}$ decay of two nuclei pairs.},sort=urca} 
\newglossaryentry{alphacen}{name=Alpha Centauri,description={one of the brightest stars in the night sky and a close binary}}
\newglossaryentry{pcygni}{name={P Cygni}, text={P Cygni},description={a hypergiant luminous blue variable with strong winds. Often referred to as a description for their line profiles showing a emission peak at the rest wavelength of the line and a blue-shifted absorption trough.}}

\newglossaryentry{teff}{name={effective temperature~}, text={effective temperature}, symbol={\ensuremath{T_\textrm{eff}}}, description={Temperature of a blackbody emitting the same total energy}, sort=teff}

\newglossaryentry{logg}{name={surface gravity~}, text={surface gravity}, symbol={\ensuremath{\textrm{log}\,g}}, description={gravity at the surface of a star}, sort=logg}
\newglossaryentry{feh}{name={metallicity~}, text={metallicity}, symbol=\textrm{[Fe/H]},description={iron abundance relative to the sun}, sort=feh}

\newglossaryentry{texp}{name={time since explosion~}, text={time since explosion}, text={time since explosion}, symbol={\ensuremath{t_{\rm exp}}},description={time since explosion (measured in days)}, sort=texp, first={time since explosion (\ensuremath{t_{\rm exp}})}}

\newglossaryentry{lmc}{name=LMC,description={Large Magellanic Cloud}, first={Large Magellanic Cloud (LMC)}, sort=lmc}
\newglossaryentry{smc}{name=SMC,description={Small Magellanic Cloud}, sort=smc}
\newglossaryentry{z}{name=\ensuremath{z},description={redshift}, sort=z}

\renewcommand{\sn}[2]{\object{SN~#1#2}}

\begin{document}

% TITLE AND AUTHORS
%-----------
\title{Very late photometry of  SN~2011fe}
\author{W.~E.~Kerzendorf\altaffilmark{1}, S.~Taubenberger\altaffilmark{2}, I.~R.~Seitenzahl\altaffilmark{3,2,4}, and A.~J.~Ruiter\altaffilmark{3, 2} } 
\email{wkerzendorf@gmail.com}

\altaffiltext{1}{Department of Astronomy and Astrophysics, University of Toronto, 50 Saint George Street, Toronto, ON M5S 3H4, Canada}
\altaffiltext{2}{Max-Planck-Institut f\"ur Astrophysik, Karl-Schwarzschild-Stra{\ss}e 1, 85748 Garching, Germany} 
\altaffiltext{3}{Research School of Astronomy and Astrophysics, Mount Stromlo Observatory, Cotter Road, Weston Creek, ACT 2611, Australia}
\altaffiltext{4}{Universit\"at W\"urzburg, Emil-Fischer-Stra{\ss}e 31,
  97074 W\"urzburg}

\begin{abstract}
The Type Ia supernova \sn{2011}{fe} is one of the closest supernovae of the past decades. Due to its proximity and low dust extinction, this object provides a very rare opportunity to study the extremely late time evolution ($>900$\,days) of thermonuclear supernovae. In this Letter, we present our photometric data of \sn{2011}{fe} taken at an unprecedented late epoch of $\approx 930$\,days with GMOS-N mounted on the Gemini North telescope ($g=23.43 \pm 0.28$, $r=24.14 \pm 0.14$, $i=23.91 \pm 0.18$, and $z=23.90 \pm 0.17$) to study the energy production and retention in the ejecta of \sn{2011}{fe}. Together with previous measurements by other groups, our result suggests that the optical supernova light curve can still be explained by the full thermalization of the decay positrons of \nucl{Co}{56}. This is in spite of  theoretical predicted effects (e.g. infrared catastrophe, positron escape, and dust) that advocate a substantial energy redistribution and/or loss via various processes that result in a more rapid dimming at these very late epochs.
\end{abstract}


\maketitle

\section{Introduction}

\sneia constitute explosive endpoints of stellar evolution, 
are major contributors to galactic chemical evolution, and as distance indicators 
are one of astronomy's most powerful cosmological tools. Despite their wide-ranging 
applications, the physical processes that lead to, result in, and sustain the transient phenomena that we know as \sneia remain relatively uncertain. 

While there is almost unanimous agreement that these events are powered by the nuclear burning of massive ($\ge1\msun$) \glspl{cowd}, there remain many open questions about the scenarios leading to the creation of these objects, the subsequent ignition, and engines that power the light curves and spectra we observe. 

Despite the uncertainty about the specifics of energy generation, the community agrees that the luminosity of \sneia is powered by the decay of radioactive nuclei produced in the explosion. The initial energy comes in the form of decay positrons, electrons, X-rays, and $\gamma$-rays, which is then reprocessed in the ejecta to UVOIR wavelengths. In particular, the $\nucl{Ni}{56} \rightarrow {\nucl{Co}{56}} \rightarrow {\nucl{Fe}{56}}$ decay chain is responsible for the majority of the energy deposition that leads to the observed luminosity. \nucl{Ni}{56} (half-life ${\sim}6\,\mathrm{d}$) has nearly fully decayed 50\,days after the explosion and the light curve is then mostly powered by the decay of \nucl{Co}{56} (half-life ${\sim}77\,\mathrm{d}$). At 300\,days the ejecta have become almost completely transparent to $\gamma$-rays. Charged decay leptons, most notably the positrons produced in $\beta^+$ decay of \nucl{Co}{56}, and low-energy X-rays can still deposit their energy and thus determine the UVOIR luminosity of the supernova. This suggests that from this time onward (at least until the internal conversion and Auger electrons produced in the decay of \nucl{Co}{57} dominate the energy injection; see e.g.~\citealt{2009MNRAS.400..531S}), the light curve should show a decline following the decay of \nucl{Co}{56}. This has been corroborated  by the relatively few normal \sneia that have been observed at these late times \citep[e.g.][]{1997A&A...328..203C, 2004A&A...428..555S, 2006AJ....132.2024L, 2007A&A...470L...1S,2009A&A...505..265L}. One should specifically mention \sn{1992}{A}, which, with observations at 926\,days past maximum, before this work held the record for the latest measurement of any spectroscopically normal \snia \citep{1997A&A...328..203C}.

The relatively strict adherence of the supernova light curves to the \nucl{Co}{56} decay at these very late epochs is puzzling, as it requires the conversion of a constant fraction of the energy produced to UVOIR band photons in the decay chain over a relatively long time. This seems to be the case despite theoretical predictions of various effects that might lead to a more rapid dimming in the observed bands and departure of the light curve decline from the \nucl{Co}{56} decay rate. Specifically, we will discuss three possible sources of deviation from an exponential light curve decline. 

As indicated previously, at late epochs (past 300\,days) the observed light curve is mainly powered by the energy deposition of decay positrons. Supernova light curves that follow the \nucl{Co}{56} decay cannot be explained by the interaction of free streaming positrons, but require being trapped through a highly tangled magnetic field \citep[see ][]{1993ApJ...405..614C, 1999ApJS..124..503M,2001ApJ...559.1019M}. In contrast, a radially combed field or no magnetic field would lead to a dimming by a factor of about five compared to a light curve following \nucl{Co}{56} at 1000\,days past explosion \citep[see Figure 1 in][]{2001ApJ...559.1019M}. Recent publications that take  into account essential near-IR corrections require almost complete trapping of positrons and thus a tangled magnetic field up to quite late epochs to explain observations \citep[e.g.][]{2007A&A...470L...1S,2009A&A...505..265L}. 

The influence of magnetic field configuration on the light curve shape leads \citet{1998ApJ...500..360R} to suggest that the light curve form can be used to confirm or rule out certain progenitor scenarios. They suggest that a tangled magnetic field, resulting in full positron trapping, stems from a Chandrasekhar mass accretion, whereas an edge-lit sub-Chandrasekhar mass scenario might produce radially combed magnetic field configuration, which enhances the escape of positrons and thus predict that this would lead to a deviation of the light curve from \nucl{Co}{56} decay. 

A second scenario leading to a departure from \nucl{Co}{56} decay, specifically in the UVOIR bands, is the so-called \gls{irc}, which predicts that the optical and near-IR light curves drop off much more rapidly after ${\sim}500\,\mathrm{d}$, even if all positrons remain trapped \citep[][]{2009A&A...505..265L}. The \gls{irc} is predicted to occur when the temperature drops below what is required to excite optical and near-IR atomic transitions ($T {\lesssim}1500$\,K), and cooling suddenly proceeds via fine structure lines emitting in the far-IR. This effect is still predicted by modern supernova radiative transfer codes \citep[e.g.][]{2009A&A...505..265L}, but is tauntingly not seen in observational data out to 786\,days after explosion for \sn{2003}{hv} \citep[e.g.][]{2004A&A...428..555S, 2009A&A...505..265L, 2014ApJ...786..134M}.

A final scenario that might lead to dimming in the UVOIR bands is the formation of dust. Unlike the IRC and positron escape scenarios the prediction by \citet{2011ApJ...736...45N} is that normal \sneia are unlikely sites of dust formation and thus predict no extinction or drastic color change of the light curve at very late times due to newly formed dust \citep[this does not necessarily extend to unusual \sneia; see][]{2013MNRAS.432.3117T}.

\sn{2011}{fe} \citep{2011Natur.480..344N, 2011Natur.480..348L} is one of the closest \sneia in the last century \citep[6.4~Mpc;][]{2011ApJ...733..124S} and is essentially unattenuated by foreground dust. The last SN~Ia that could have been observed in similarly exquisite detail as \sn{2011}{fe} (if today's technology had been available) was \sn{1972}{E} \citep{1972ApJ...177L..59L}. This allows for unprecedented observations of this object out to very late phases and presents us a rare opportunity to test theoretical predictions about the light curve and spectral evolution. 

In this work, we present photometric observations of SN~2011fe at the extremely late epoch of $\approx 930$\,days past maximum. In Section~\ref{sec:obs_analysis}, we give a description of the observations and subsequent data reduction. Section~\ref{sec:discussion} is devoted to discussing the observations when compared to theoretical predictions. We present our conclusions and discuss possible future work in Section~\ref{sec:conclusion_future}. 

\section{Observations \& Analysis}
\label{sec:obs_analysis}

\begin{figure*}[ht!]
\includegraphics[width=0.47\textwidth]{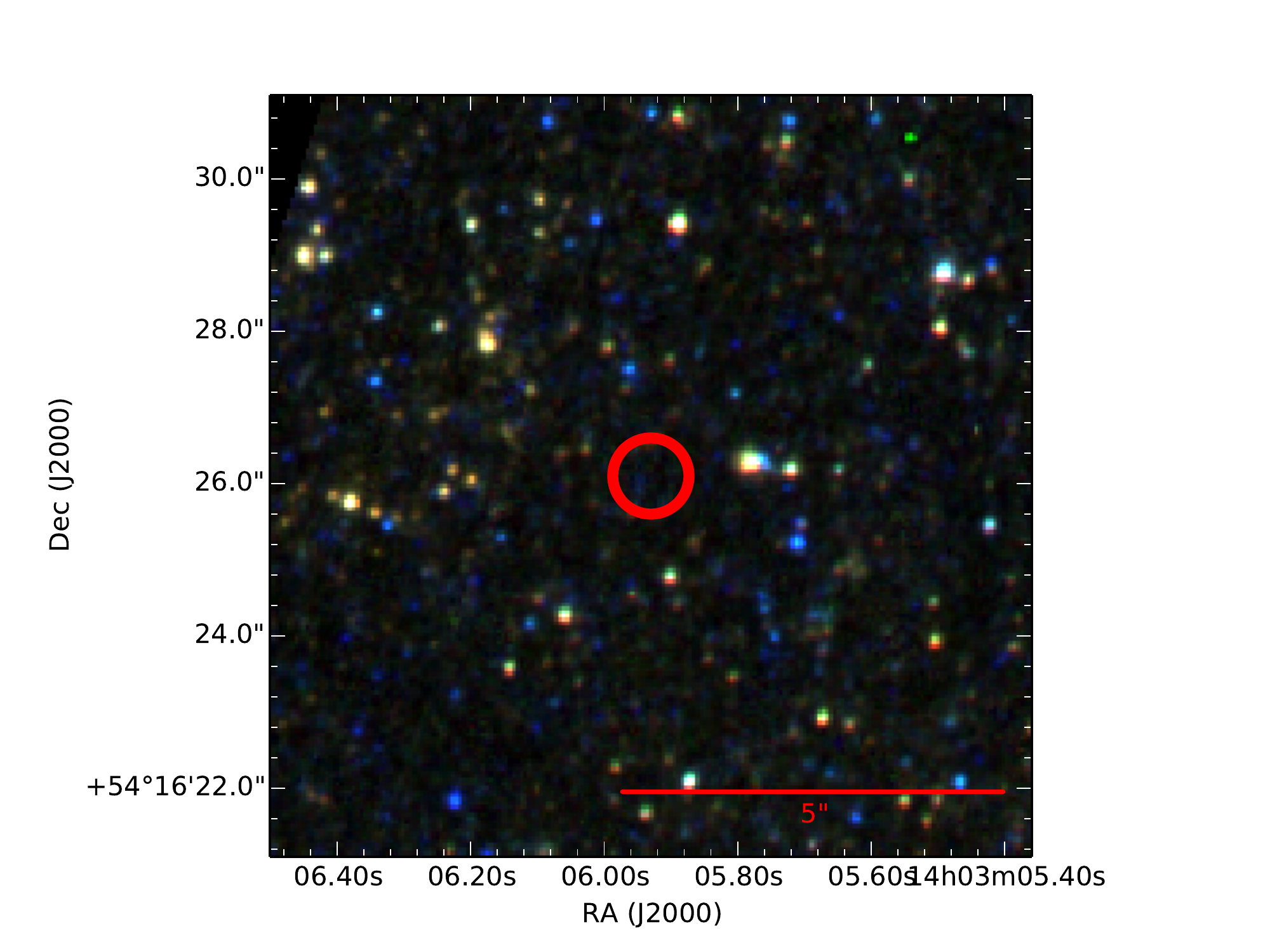}
\includegraphics[width=0.47\textwidth]{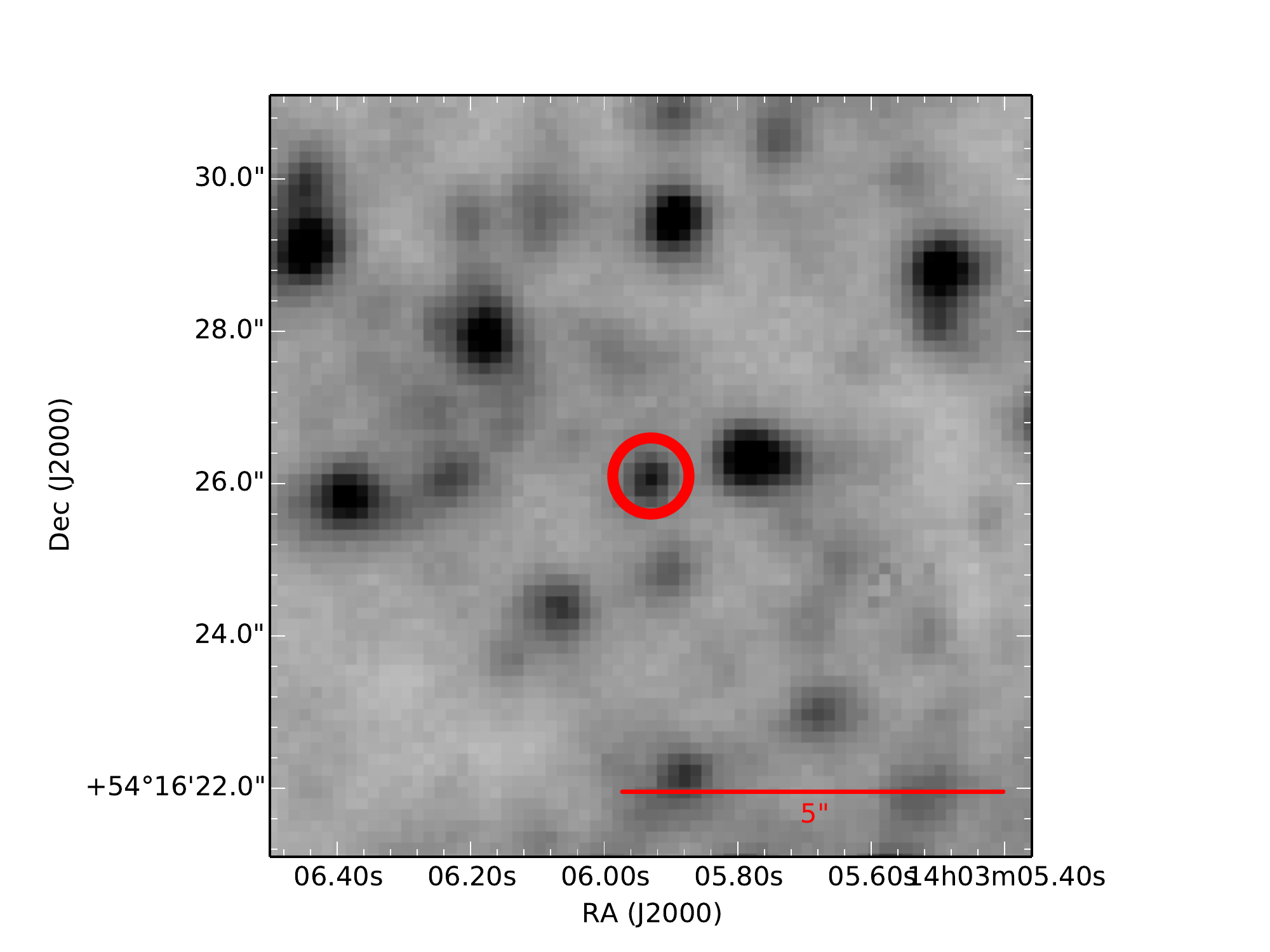}
\caption{\textbf{Left side:} A color composite image from observations taken by the \gls{acs} under program GO9490 (PI: K. Kuntz) before the explosion of SN~2011fe. The circle diameter is $1^{\prime\prime}$. \textbf{Right side:} GMOS g-band image taken with our program (GN-2014A-Q-24) $+909$\,days past maximum.}
\label{fig:sn2011fe_detection}
\end{figure*}

\ctable[
caption=Photometry of SN~2011fe,
label=tab:observations,
pos=tb!,
star
]
{lcccccc}{
\tnote{Assuming $\textrm{B}_\textrm{max}$ at MJD 55814.51 {\citep{2013A&A...554A..27P}}}
}{\FL
Date & $t - t(\textrm{B}_\textrm{max})$\tmark[a] & Filter & mean airmass & seeing & $t_\textrm{exposure}$  & magnitude \NN
YYYY MM DD & d &  &  &  $^{\prime\prime}$ & s & mag \ML
 2014 03 07 & 909.1 &  g & 1.3 & 0.57 & $180 \times 5$ & $23.43 \pm 0.28$\\
 2014 03 27 & 929.1&  r & 1.4 & 0.53 & $180 \times 5$ & $24.14 \pm 0.14$\\
 2014 03 28 & 929.9& i  & 1.4 & 0.60 & $180 \times 5$ & $23.91 \pm 0.18$\\
 2014 03 28 & 929.9& z & 1.3 & 0.57 & $180 \times 10$ & $23.90 \pm 0.17$
\LL}

We obtained optical photometry in the $g$, $r$, $i$ and $z$ bands using  \gls{gmosn} mounted on the Gemini North
telescope located at Mauna Kea (program GN-2014A-Q-24). The data were taken on the nights of 2014 March 7, 27, and 28 (see Table~\ref{tab:observations}) under photometric conditions. The data were then pre-reduced with the \textsc{geminiutil}\footnote{{\url{http://github.com/geminiutil/geminiutil}}} package following standard procedures. After careful inspection of the \glsdesc{wcs}, the images were aligned and combined using \gls{swarp}. In a final step, we adjusted the astrometric calibration to match \gls{acs} observations (see Figure~\ref{fig:sn2011fe_detection}). We undertook the same operations on SDSS-calibrated stars \citep{2011ApJS..193...29A} in the field and then used those for the calibration. Finally, we also observed standard fields during the same nights (DLS 1359-11, PG1633+099) to make an additional cross-check for calibration with SDSS, showing consistency in each filter.

Subsequently, we performed our measurements using \gls{psf} photometry with the \textsc{snoopy} package. This package is a compilation of \gls{iraf} tasks optimized for \gls{sn} photometry, developed by F. Patat and E. Cappellaro.
\textsc{snoopy} constructs
the \gls{psf} by selecting several clean unblended stars and then performs \gls{psf} photometry on the \gls{sn} itself. The instrumental SN magnitudes were finally calibrated to the Sloan photometric system \citep{1996AJ....111.1748F} using tabulated atmospheric extinction coefficients and the nightly zero points derived from our standard-field observations. To get a better estimate of the uncertainty of the photometric measurement, \textsc{snoopy} uses an artificial star experiment.

\begin{figure}[ht!]
\begin{center}
\includegraphics[width=0.5\textwidth]{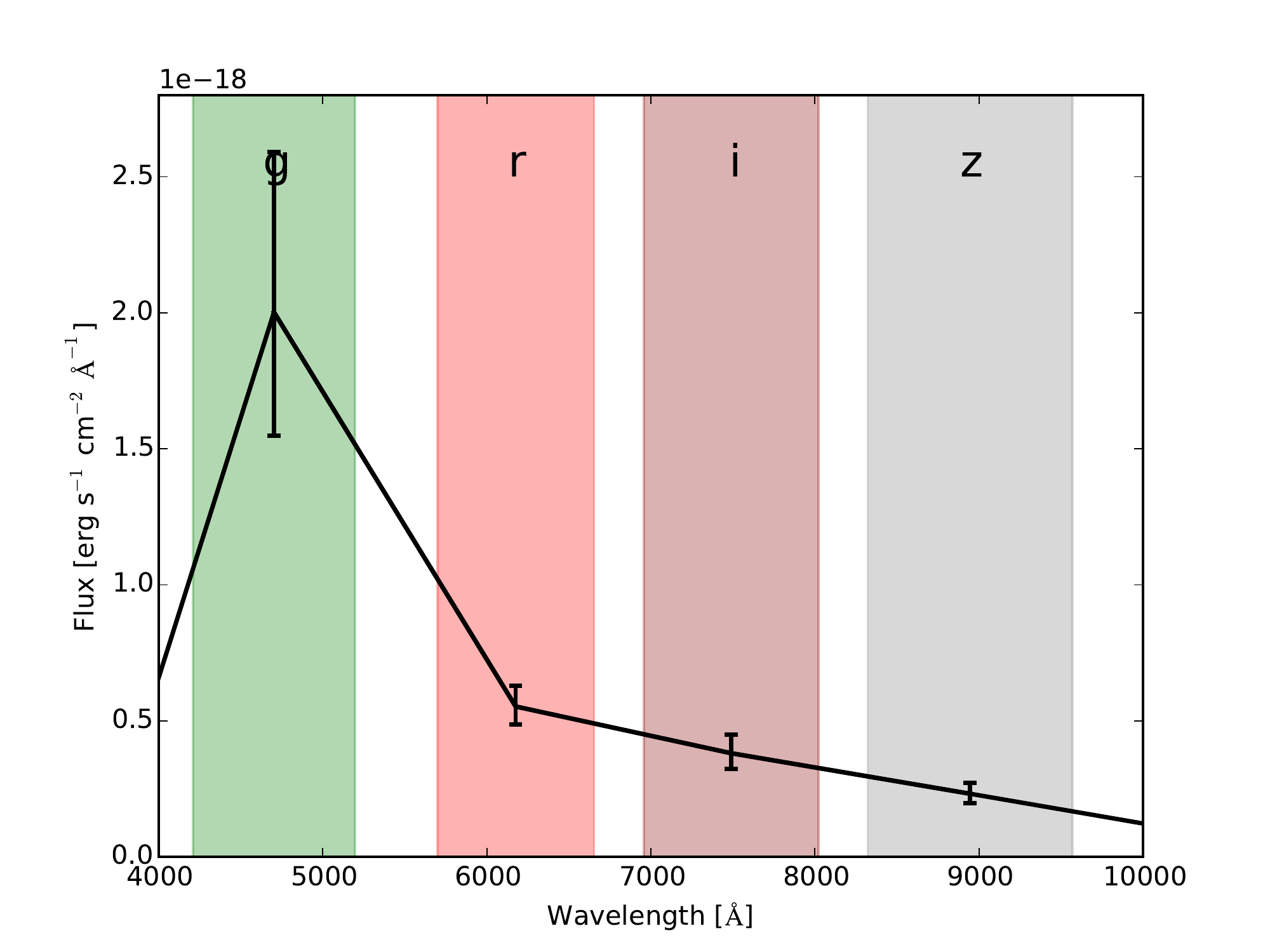}
\end{center}
\caption{The SED constructed from our Gemini photometry at 930\,days. The $g$-band photometry was taken 21 days 
earlier, but has been scaled accordingly using the \citet{2012ApJ...750L..19R} luminosity evolution.}
\label{fig:sn2011fe_sed}
\end{figure}

Finally, we combined our photometric measurements into a pseudo bolometric luminosity. The $g$-band measurement was taken 21\,days earlier than the other bands and hence we applied a dimming correction of 0.15 mag based on a theoretical bolometric light curve model of \citet{2012ApJ...750L..19R}. First, we generated a spectral energy distribution (SED; see Figure~\ref{fig:sn2011fe_sed}) using the weighted mean wavelengths of the bands as the supporting points. We repeated the same procedure for the photometric data of \citet{2013CoSka..43...94T}, which unlike our photometry uses Bessell \textit{BVR} filters. We used linear interpolation for each band in their light curves to obtain a set of \textit{BVR} magnitudes at 400\,days and 550\,days. The SEDs generated from the measurements  presented in this work and from \citet{2013CoSka..43...94T} were then integrated between 4000\AA\ and 8000\AA\ (this wavelength range was chosen as it is covered by both filter systems), assuming a linear flux distribution between the supporting points (see Figure~\ref{fig:sn2011fe_sed}). Furthermore, we assumed that, at edges of the magnitude sets, our flux distribution is zero (e.g., at the blue edge of the $B$ filter and at the red edge of the $R$ filter; same for $g$-band and $z$-band; see Figure~\ref{fig:sn2011fe_sed}). The final pseudo bolometric uncertainties were determined using Monte Carlo techniques. The result can be seen in Figure~\ref{fig:luminosity_decay}. 

\section{Discussion}
\label{sec:discussion}

\begin{figure*}[ht!]
\includegraphics[width=1\textwidth]{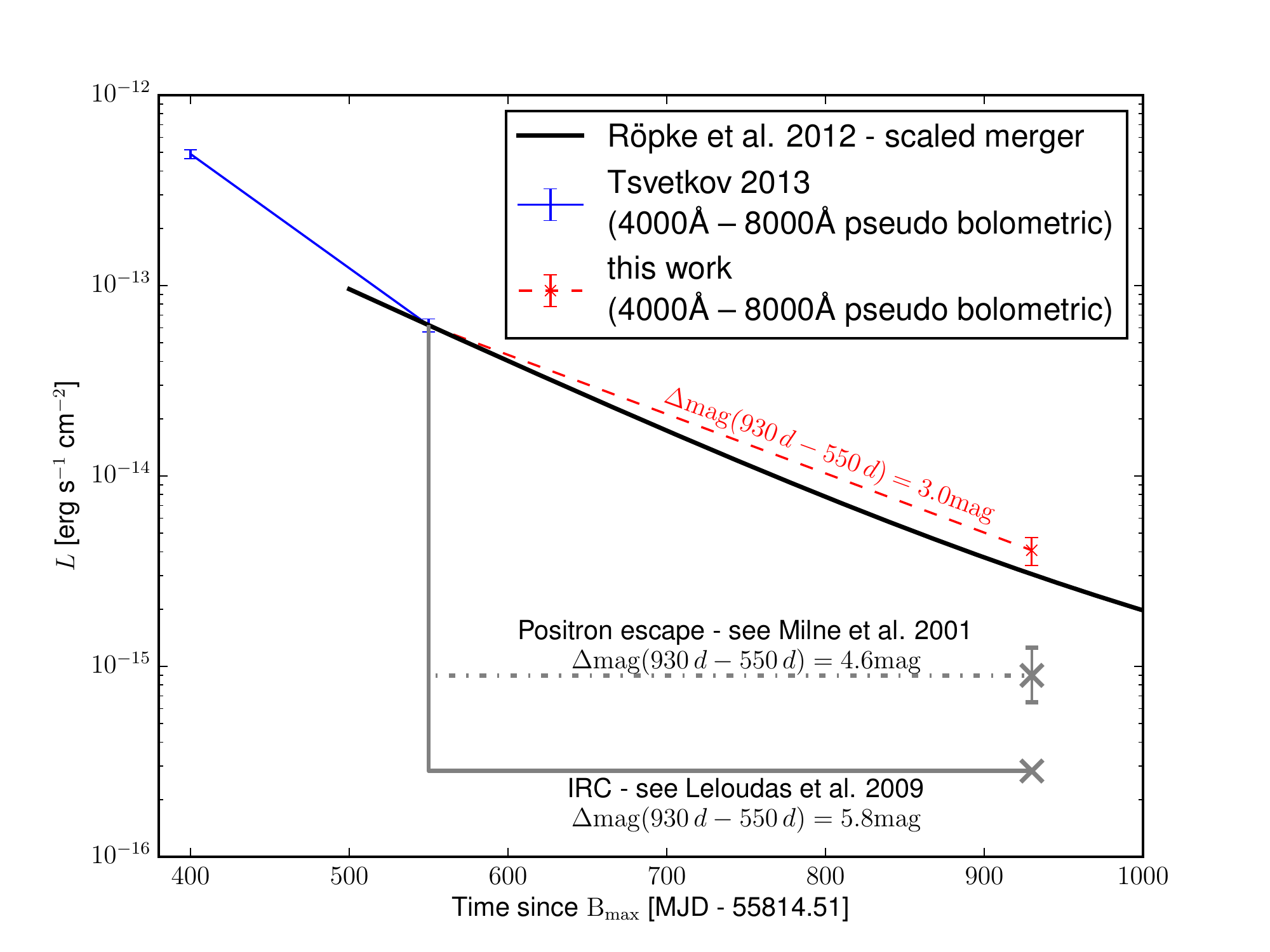}

\caption{Late-time pseudo bolometric light curve of SN~2011fe (integrated between 4000\AA\ and 8000\AA). We compare pseudo bolometric measurements composed from \textit{BVR} data of \citet{2013CoSka..43...94T} to our pseudo bolometric measurements. Phases are reported relative to B-band maximum light 
\citep[MJD 55814.51;][]{2013A&A...554A..27P}. We added the \gls{cowd} merger bolometric light curve from \citet{2012ApJ...750L..19R} assuming a rise time of 20.8\,days to $\textrm{B}_\textrm{max}$ \citep[][]{2012ApJ...747L..10P} and scaled to match the pseudo bolometric luminosity at 550\,days. Finally, we added the decline between 550\,days and 930\,days predicted by scenarios with an IRC or positron escape.}
\label{fig:luminosity_decay}
\end{figure*}

We have acquired optical photometry in the period between 909\,days and 930\,days past maximum and are comparing these data to earlier photometry by \citet{2013CoSka..43...94T} in Figure~\ref{fig:luminosity_decay}. This comparison shows a decline that is broadly consistent (at $1.5\sigma$) with \nucl{Co}{56} decay as seen by the scaled (bolometric) light curve taken from the merger model of \citet{2012ApJ...750L..19R}. 

The comparison is complicated by the fact that the data from \citet{2013CoSka..43...94T} are in the Johnson-Cousins photometric system \citep[see][]{1998A&A...333..231B} and our data are in filters that are similar to SDSS \citep{1996AJ....111.1748F}. Calculating transformations from one system to the other requires detailed spectral information, which was not obtained with our program. We alleviated this problem by reconstructing an SED out of the filter measurements and then integrating these over a wavelength range that is covered by both filter sets (4000\AA\ -- 8000\AA; see Section~\ref{sec:obs_analysis} for details; see Figure~\ref{fig:luminosity_decay}). 

For a comparison with the observed pseudo bolometric light curve we opt for an analytic theoretical light curve model of a violent merger \citep{2012ApJ...750L..19R}. This bolometric light curve considers gamma rays as free streaming and assumes complete deposition and instantaneous thermalization of X-rays and positron and electron kinetic energies produced in radioactive decays, which is a good approximation at phases later than 500\,days. Under these assumptions, the model closely follows the slope of \nucl{Co}{56} decay, and is generic for \sneia. In fact, the model light curve for a Chandrasekhar-mass explosion that was also presented by \citet{2012ApJ...750L..19R} is nearly identical at the epochs under consideration. This also means that the presented photometry cannot be used the distinguish between progenitor models. For comparison, we have scaled the bolometric light curves to match the pseudo-bolometric data point generated from the \citet{2013CoSka..43...94T} data at 550\,days. 

Figure~\ref{fig:luminosity_decay} shows that the observed pseudo bolometric light curve is broadly consistent with the scaled bolometric model by \citet{2012ApJ...750L..19R}. The predicted pseudo bolometric luminosity decline in the case of an \gls{irc} was calculated from synthetic \textit{BVR} magnitudes \citep[see Figure 9 in][]{2009A&A...505..265L} in the same way as for the observed data. For the luminosity decline following an enhanced positron escape there only exist true bolometric models, which cover a certain range in $\Delta L$ as indicated in Figure~\ref{fig:luminosity_decay}. The reader should note that this is not a $1\sigma$ error bar, but a range of models for different ionization fractions presented in \citet{2001ApJ...559.1019M}. Finally, both models were anchored at 550\,days  and Figure~\ref{fig:luminosity_decay} shows the predicted declines between 550\,days and 930\,days. 

There is a discrepancy between 400\,days and 550\,days, where the observations dim more rapidly (and also more rapidly than \nucl{Co}{56} decay), but we speculate that the seemingly underluminous data point at 550\,days is caused by an increasing amount of flux shifted out of the optical bands into the infrared, which is corroborated by the behaviour of \sn{2003}{hv} \citep[see Figure 8]{2009A&A...505..265L}. After 550\,days, \sn{2003}{hv} shows a reversed trend \citep{2009A&A...505..265L}.

The most straightforward explanation for the observed pseudo-bolometric evolution is thus a true bolometric light curve decline that indeed follows the radioactive decay of \nucl{Co}{56} and \nucl{Co}{57}, combined with a temporally slightly varying bolometric correction. This explanation suggests a fully thermalized positron kinetic energy and no IRC. Our data, however, cannot exclude a finely tuned model that, while having a true bolometric evolution that deviates from the radioactive decay slope, still almost perfectly compensates this by a change in the bolometric correction, resulting in a seemingly good fit in the optical.

One such possibility would be a scenario in which a rapid dimming owing to positron escape or an IRC is just compensated by a light echo. In this case, the observed spectrum of SN 2011fe should be similar to the spectrum at maximum light \citep[e.g.][]{1994ApJ...434L..19S}. We thus compare magnitudes of \sn{2011}{fe}, calculated using \gls{synphot} maximum light spectrum ($t - t_\textrm{max} = -0.03$\,days) of \citet{2013A&A...554A..27P} ($g=9.89$, $r=10.05$, $i=10.63$ and $z=10.77$) with our own measurements at 930\,days (see Table~\ref{tab:observations}). The comparison shows that at 930\,days \sn{2011}{fe} has a bluer $g - r$ color [$(g-r)_{930\,\textrm{d}} = -0.51$ versus $(g-r)_\textrm{max}$ = -0.16] and at the same time a redder $r-i$ color [$(r-i)_{930\,\textrm{d}} = +0.23$ versus $(r-i)_\textrm{max}$ = -0.58] than at maximum light. The very different colors suggests that the majority of the light is originating in the ejecta and there is no major contribution of scattered light from earlier epochs.

\section{Conclusion and Future Work}
\label{sec:conclusion_future}

With the observations of \sn{2011}{fe} shown in this work, we have presented the latest photometric data for any \snia if one discounts the spectroscopically peculiar \sn{1991}{T}, which could be observed even at 2570 days due to a strong light echo produced by dust between the \gls{sn} and the observer \citep{1999ApJ...523..585S}. Combining previous photometry from \citet{2013CoSka..43...94T} with our work shows \sn{2011}{fe} to be consistent with full thermalization of all \nucl{Co}{56} decay positrons until at least $\approx 930$\,days past maximum. There is no evidence for various dimming effects that have been suggested by theory. In fact the dimmed models are more than $4\sigma$ outliers compared to the  datapoint luminosity at 930\,days. 

The current data indicate full trapping of positrons (significant positron escape would require the luminosity to be dimmer by a factor of approximately five at 930\,days when compared to \nucl{Co}{56} decay; see Figure~\ref{fig:luminosity_decay}) , which, combined with the predictions by \citet{1998ApJ...500..360R}, favors the accreting Chandrasekhar mass \gls{cowd} over the sub-Chandrasekhar mass edge-lit \gls{cowd} \citep[the reader should note that only those two models were compared in][]{1998ApJ...500..360R}. This emphasizes the importance of studying the magnetic field configurations in various competing scenarios \citep[particularly the ones not mentioned in][]{1998ApJ...500..360R}.
Furthermore, the observations are not compatible with the predicted \gls{irc} that suggests cooling only via far-IR lines resulting in a complete drop of the optical and near-IR luminosities. This might indicate that the ejecta are currently still above a critical temperature of $T \approx 1500$\,K. Finally, we can rule out the formation of large amounts of dust on the basis of both the current brightness (Figure~\ref{fig:luminosity_decay}) and spectral energy distribution (Figure~\ref{fig:sn2011fe_sed}) of the supernova. This is consistent with the predictions that normal \sneia do not produce dust in significant amounts \citep[e.g.][]{2011ApJ...736...45N}. A final caveat that might call the previous conclusions into question is that the measurements could be significantly contaminated by a light echo. However, as discussed in Section~\ref{sec:discussion}, the colors at the current epoch are significantly different from those at maximum light, allowing at most a very small contribution from a light echo and re-affirming our conclusions.

The fact that \sn{2011}{fe} is still relatively bright provides a unique opportunity to study the very late phase behavior of this \gls{sn} in unprecedented detail. We aim to continue this project by observing \sn{2011}{fe} at future epochs (${\sim}1500$ days past maximum) in optical and near-IR bands, allowing us to measure a quasi-bolometric luminosity evolution to determine even more precisely the energy deposition in the \gls{sn} ejecta at such late phases, which can then be more directly confronted with theoretical predictions. Finally, we hope that our results encourage the community to continue observing \sn{2011}{fe} with a variety of different techniques. 

\acknowledgements

This research made use of Astropy, a community-developed core Python package for Astronomy \citep{2013A&A...558A..33A}.
I.R.S. and A.J.R. acknowledge funding from ARC Laureate Grant FL0992131. S.T. acknowledges support by the Transregional Collaborative Research Centre TRR 33 of the German Research Foundation.
We would like to thank Marten van Kerkwijk for many illuminating discussions and Bruno Leibundgut for useful discussions on light echoes. We also thank the Gemini team for the help they provided using the telescope, in particular Katherine Roth, who went above and beyond her call of duty to provide help with this program. Finally, we thank the anonymous referee for helpful suggestions to improve the manuscript. 
Based on observations obtained at the Gemini Observatory, which is operated by the 
Association of Universities for Research in Astronomy, Inc., under a cooperative agreement 
with the NSF on behalf of the Gemini partnership: the National Science Foundation 
(United States), the National Research Council (Canada), CONICYT (Chile), the Australian 
Research Council (Australia), Minist\'{e}rio da Ci\^{e}ncia, Tecnologia e Inova\c{c}\~{a}o 
(Brazil) and Ministerio de Ciencia, Tecnolog\'{i}a e Innovaci\'{o}n Productiva (Argentina).

\textit{Facility:} \facility{Gemini:Gillett (GMOS)}

\bibliographystyle{apj}
\bibliography{sn2011fe_phot1}

\end{document}